\documentclass[english,conference,ruled]{IEEEtran}
\usepackage[T1]{fontenc}
\usepackage[latin9]{inputenc}
\usepackage{geometry}
\usepackage[active]{srcltx}
\usepackage{color}
\usepackage{babel}
\usepackage{mathtools}
\usepackage{algorithm2e}
\usepackage{amsmath}
\usepackage{amsthm}
\usepackage{amssymb}
\usepackage{graphicx}
\usepackage{esint}
\usepackage[unicode=true,pdfusetitle,
 bookmarks=true,bookmarksnumbered=false,bookmarksopen=false,
 breaklinks=false,pdfborder={0 0 1},backref=false,colorlinks=true]
 {hyperref}
\hypersetup{
 pdfborderstyle=}

\makeatletter
\theoremstyle{plain}
\newtheorem{thm}{\protect\theoremname}

\usepackage{babel}
\newcommand{\nosemic}{\renewcommand{\@endalgocfline}{\relax}}
\newcommand{\dosemic}{\renewcommand{\@endalgocfline}{\algocf@endline}}
\usepackage{cite}
\usepackage{subfigure}
\newtheorem{lemma}{Lemma}\newtheorem{remark}{Remark}\newtheorem{definition}{Definition}

\providecommand{\theoremname}{Theorem}

\makeatother

\providecommand{\theoremname}{Theorem}

\begin{document}
\newgeometry{top=1in,bottom=0.75in,right=0.75in,left=0.75in} \IEEEoverridecommandlockouts 
\title{Extension of Full and Reduced Order Observers for Image-based Depth
Estimation using Concurrent Learning}
\author{Ghananeel Rotithor, Daniel Trombetta, Rushikesh Kamalapurkar, Ashwin
Dani\thanks{Ghananeel Rotithor, Daniel Trombetta and Ashwin P. Dani are with the
Department of Electrical and Computer Engineering at University of
Connecticut, Storrs, CT 06269. Email: \{ghananeel.rotithor; daniel.trombetta;
ashwin.dani\}@uconn.edu. Rushikesh Kamalapurkar is with the Mechanical
and Aerospace Engineering Department at the Oklahoma State University,
OK. Email: rushikesh.kamalapurkar@okstate.edu}}
\maketitle
\begin{abstract}
In this paper concurrent learning (CL)-based full and reduced order
observers for a perspective dynamical system (PDS) are developed.
The PDS is a widely used model for estimating the depth of a feature
point from a sequence of camera images. Building on the current progress
of CL for parameter estimation in adaptive control, a state observer
is developed for the PDS model where the inverse depth appears as
a time-varying parameter in the dynamics. The data recorded over a
sliding time window in the near past is used in the CL term to design
the full and the reduced order state observers. A Lyapunov-based stability
analysis is carried out to prove the uniformly ultimately bounded
(UUB) stability of the developed observers. Simulation results are
presented to validate the accuracy and convergence of the developed
observers in terms of convergence time, root mean square error (RMSE)
and mean absolute percentage error (MAPE) metrics. Real world depth
estimation experiments are performed to demonstrate the performance
of the observers using aforementioned metrics on a 7-DoF manipulator
with an eye-in-hand configuration. 
\end{abstract}

\section{Introduction}

Estimating the 3D coordinates of feature points using observations
from a sequence of camera images is referred to as the Structure from
Motion (SfM) problem in computer vision literature. The 3D coordinates
of feature points can be estimated by estimating the depth of the
features. The estimated 3D coordinates of feature points or structure
information can be used in a variety of automatic control, autonomy,
and intelligent control applications. Existing solutions to this problem
include offline \cite{ma2012invitation} and online \cite{spica2014active,Dani2012a,dani2012globally,matthies1989kalman,chiuso2002structure,jankovic1995visually,karagiannis2005new,Dixon2003f,dahl2010observer,chen2004state,DeLuca2008,keshavan2018robust,gans2009image,hatanaka2012cooperative,Hu2008}
methods. The focus of this paper is on online methods where the problem
is formulated as a state estimation problem of a perspective dynamical
system (PDS). The PDS is a class of nonlinear system that uses inverse
depth parameterization, which is widely used in observer-based methods,
and simultaneous localization and mapping (SLAM) \cite{Yang2015a}.

Online methods often rely on the use of an Extended Kalman Filter
(EKF) \cite{matthies1989kalman,chiuso2002structure}. In comparison
to EKF-based approaches, nonlinear observers are developed for SfM
with analytical proofs of stability. Under the assumption that the
camera motion is known, continuous and discontinuous observers are
developed to estimate the depth which can then be used to estimate
the range to the object. A high-gain observer called the identifier-based
observer is presented for range estimation in \cite{jankovic1995visually}.
A semi-globally asymptotically stable reduced-order observer is presented
in \cite{karagiannis2005new} to estimate the range based on immersion
and invariance (I\&I) methodology, which is extended to the range
and orientation identification observer design in \cite{sassano2010observer}.
A continuous observer, which guarantees asymptotic range estimation,
is presented in \cite{Dixon2003f} under the assumption that camera
motion is known. In \cite{dahl2010observer}, an asymptotically converging
nonlinear observer is developed based on Lyapunov's indirect method.
In \cite{chen2004state}, a discontinuous sliding-mode observer is
developed which guarantees exponential convergence of the estimation
error. In \cite{DeLuca2008}, a nonlinear observer is developed that
achieves local exponential convergence of estimation error. A range
observer design based on nonlinear contraction and synchronization
theory is presented in \cite{grave2014new}. In \cite{dani2012globally},
a globally exponentially stable observer is designed for the PDS.
Extensions of these observers for PDS with moving objects are presented
in \cite{DaniT.A.a,Dani2010b}. All these observers require persistence
of excitation (PE) condition to be satisfied by the camera motion
to achieve the convergence of the estimation error.

Drawing parallels to the adaptive control/observer design, in the
PDS the inverse depth appears as a parameter in the dynamics of image-plane
coordinates where the parameter is time-varying with known dynamics
associated with it. Concurrent Learning (CL) is used in adaptive control
for parameter estimation, where the knowledge of past trajectory data
is leveraged to estimate the constant parameter and achieve state
tracking \cite{chowdhary2013concurrent,kamalapurkar2017concurrent}.
CL has also been used for target tracking applications in \cite{parikh2018target}
and for target size estimation in \cite{fairfax2018concurrent}. The
use of CL relaxes the PE condition to a finite excitation condition,
which depends upon the rank of the regressor matrix \cite{chowdhary2013concurrent}.

Inspired by the recent advances in CL in adaptive control, full and
reduced-order depth observers are proposed in the paper that builds
on the work in \cite{dani2012globally}. The observer design guarantees
the boundedness of the depth estimation error even when the PE condition
is not satisfied by the camera motion in a time window given that
finite excitation is present. The observer can be used to estimate
the feature point depth to a desired accuracy. Two cases are analyzed
for the convergence and stability of the observer design. The fist
case is when the camera motions satisfy the PE condition and the second
case is when the camera motions do not satisfy the PE condition. A
Lyapunov stability analysis is carried out for the switched observer
error system (i.e., when PE is satisfied and when it is not) using
multiple Lyapunov functions \cite{vu2007input}. Although compared
to the existing depth/range observers in literature, the observers
in this paper cannot achieve asymptotic or exponential depth estimation
convergence, the observer can achieve finite estimation errors in
practical scenarios when the existing observers may not yield finite
estimation error. Practical examples include the motion of the camera
along the projected ray while grasping an object or when an aerial
robot is moving in the $Z$ direction during landing and takeoff or
robots moving in the direction of view while docking. For these cases
the camera motion will not satisfy the PE condition for certain time
window. Compared to the recent work in \cite{Rotithor2019,rotithor2019reduced},
a rigorous stability analysis and detailed simulation and experimental
evaluations are presented in this paper. A history stack update procedure
based on the Lyapunov analysis, which stores the camera motion and
feature point data, is presented. The results of a performance evaluation
of the CL-based full order and reduced order observers with a benchmark
observer is presented using real world experiments conducted on camera
mounted in the hand of 7 DoF Baxter robot.

\section{Perspective Camera Motion Model\label{sec:System-Modeling}}

The movement of a perspective camera capturing a scene results in
the change of image plane coordinates of a feature point belonging
to a static object. Let $\bar{m}(t)=\left[X(t)\:Y(t)\:Z(t)\right]^{T}\in\mathbb{R}^{3}$
and $m_{n}(t)=\left[\frac{X(t)}{Z(t)}\:\frac{Y(t)}{Z(t)}\:1\right]\in\mathbb{R}^{3}$
be the Euclidean and normalized Euclidean coordinates of a feature
point belonging to a static object captured by a moving camera in
the camera reference frame $\mathcal{F}_{C}$ with known camera velocities.
To estimate the depth, define an auxiliary vector $\left[x(t)\:y(t)\:\chi(t)\right]^{T}\in\mathcal{Y}$
such that $\mathcal{Y}\subset\mathbb{R}^{3}$ is a closed and bounded
set where $x(t)=\frac{X(t)}{Z(t)},y(t)=\frac{Y(t)}{Z(t)},\chi(t)=\frac{1}{Z(t)}$.
Let $s(t)=\left[x\left(t\right)\:y\left(t\right)\right]^{T}$ be the
state associated with an image plane feature point with two components
and $\chi(t)$ be the inverse depth of the feature point.

\begin{remark}\label{rem:rem1}The state variables $x$ and $y$
are image plane coordinates of a feature point whose pixel coordinates
are bounded by the resolution of the camera. As a result, the state
variables $x(t)$ and $y(t)$ are bounded by known constants $\underline{x}\leq x(t)\leq\overline{x}$
and $\underline{y}\leq y(t)\leq\overline{y}$. The Euclidean distance
$Z$(t) between the camera and the feature point can be lower bounded
by the focal length of the camera $\lambda$ measured in meters and
is not assumed to be upper bounded. Therefore, the inverse depth $\chi(t)$
can be upper and lower bounded as in \cite{dani2012globally,DeLuca2008}
using the constants $0<\underline{y_{3}}<\chi\leq\frac{1}{\lambda}$.\end{remark}
\textbf{Assumption }$\textbf{1}$:\label{Assumption-:-assumption1-1}
The depth of the feature point $Z(t)$ is invertible in the compact
set $\mathcal{Y}$.\\
 The feature point dynamics can be written as a function of the linear
and angular velocities as 
\begin{align}
\dot{s} & =f_{m}(s,\omega)+\Omega^{T}(s,v)\chi\label{eq:SysDyn}\\
\dot{\chi} & =f_{u}\left(s,\chi,u\right)\label{eq:SysDyn2}
\end{align}
where $v\left(t\right)=\left[v_{X}(t)\:v_{Y}(t)\:v_{Z}(t)\right]^{T}\in\mathcal{V}$
, $\omega\left(t\right)=\left[\omega_{X}(t)\:\omega_{Y}(t)\:\omega_{Z}(t)\right]^{T}\in\mathcal{W}$
are the linear velocities in $\frac{\mathrm{m}}{\mathrm{s}}$ and
angular velocities in $\frac{\mathrm{rad}}{\mathrm{s}}$ of the camera
in the body frame and $u\left(t\right)=\left[v^{T}(t)\:\omega{}^{T}(t)\right]^{T}$.
The sets $\mathcal{V}$ and $\mathcal{W}$ are bounded such that $\mathcal{V\subset\mathbb{R}}^{3}$
and $\mathcal{W\subset\mathbb{R}}^{3}$. In (\ref{eq:SysDyn}), $f_{m}(s,\omega)\in\mathbb{R}^{2}$
and $\Omega(s,v)\in\mathbb{R}^{1\times2}$ are functions of measurable
quantities or known quantities. The state derivative $\dot{s}\in\mathbb{R}^{2}$
is not measurable in this case and can only be estimated. Individually,
$f_{m}(s,\omega)$, $\text{\ensuremath{\Omega(s,v)}}$, and $f_{u}(s,\chi,u)$
are defined as 
\begin{gather}
f_{m}(s,\omega)=\left[\begin{array}{c}
\begin{array}{ccc}
xy & -\left(1+x^{2}\right) & y\\
1+y^{2} & -xy & -x
\end{array}\end{array}\right]\omega\nonumber \\
\Omega(s,v)=\left[\begin{array}{c}
\begin{array}{cc}
xv_{Z}-v_{x} & yv_{Z}-v_{Y}\end{array}\end{array}\right]\nonumber \\
f_{u}(s,\chi,u)=v_{Z}\chi^{2}+\left(y\omega_{X}-x\omega_{Y}\right)\chi\label{eq:Dynamics}
\end{gather}
\textbf{Problem Definition: }Given the measurements of feature points
in the image plane $s\left(t\right)$, the linear and angular velocity
of the camera $u\left(t\right)$ and the linear acceleration of the
camera $\dot{v}\left(t\right)$ in the camera reference frame, it
is desired to estimate the inverse depth of the feature point $\chi\left(t\right)$
using the dynamics in (\ref{eq:SysDyn})-(\ref{eq:SysDyn2}). To this
end, full order and reduced order depth observers are designed in
Section \ref{sec:Concurrent-Learning-based-Observ} and Section \ref{sec:Reduced-Order-Observer}
using CL\textbf{.}

\textbf{Assumption }$\textbf{2}$:\label{Assumption-:-assumption1}
The camera velocities are bounded and the linear velocities are $\mathcal{C}^{1}$
with respect to time.

\section{CL-based Full Order Observer\label{sec:Concurrent-Learning-based-Observ}}

The depth estimation schemes in the existing literature require a
strong observability condition called Persistence of Excitation (PE).
For such observers, the estimation error converges to zero only if
the PE condition is satisfied. The PE condition is satisfied if there
exist constants $T_{0},\rho\in\mathbb{R}^{+}$ such that 
\begin{alignat}{1}
\int_{t}^{t+T_{0}}\Omega(s(\tau),v(\tau))\Omega^{T}(s(\tau),v(\tau))d\tau & \geq\rho>0,\forall t>t_{0}.\label{eq:PECond}
\end{alignat}
CL based parameter estimation techniques use a history stack of recorded
data generated by the dynamical system to make updates to the parameter
estimation scheme. CL is based on the premise that even if the PE
condition can not be guaranteed, input can be exciting over a finite
interval of time. For the full order CL observer, the history stack
is a tuple $\mathcal{H=}\left\{ \left(\dot{\bar{s}}_{j},s_{j},u_{j}\right)\right\} _{j=1}^{M}$
containing the past data points up to the index $M-1$ chosen by the
algorithm proposed in Section \ref{sec:History-Stack-Update} where
$M$ is the index of the data point at the current time instant. Let
$\left\{ t_{j}\right\} _{j=1}^{M-1}$ denote the corresponding time
instances at which the $j^{\mathrm{th}}$ entry in the history stack
is recorded. Then by the definition of the history stack $\dot{\bar{s}}_{j}\coloneqq\dot{\bar{s}}\left(t_{j}\right),s_{j}\coloneqq s\left(t_{j}\right),u_{j}\coloneqq u\left(t_{j}\right)\:\forall j=1,\cdots,M-1$.

\textbf{Assumption }$\textbf{3}$: The term $\dot{\bar{s}}_{j}$ is
the approximation of $\dot{s}_{j}$ using computed numerically such
that $\Vert\dot{\bar{s}}-\dot{s}\Vert<\bar{d}$ and $\bar{d}\in[0,\infty)$
is an unknown constant.

The estimates of $s,\chi$ are denoted by $\hat{s},\hat{\chi}$ respectively
and the state and depth estimation errors as $z=\chi-\hat{\chi},$
and $\xi=s-\hat{s}$. Using the dynamics in (\ref{eq:SysDyn}) and
(\ref{eq:SysDyn2}), the observer for estimating the state and the
depth is designed as follows. 
\begin{align}
\dot{\hat{s}}= & f_{m}(s,\omega)+\Omega^{T}(s,v)\hat{\chi}+H\xi\label{eq:StateEst}\\
\dot{\hat{\chi}}= & f_{u}(s,\hat{\chi},u)+\Gamma\Omega(s,v)\xi+K_{CL}\Gamma\sum_{j=1}^{M}\Omega(s_{j},v_{j})(\dot{\bar{s}}_{j}\nonumber \\
 & -f_{m}(s_{j},\omega_{j})-\Omega^{T}(s_{j},v_{j})\hat{\chi})\label{eq:DepthEst}
\end{align}
where $H\in\mathbb{R}^{2\times2}$ is positive definite diagonal gain
matrix, $\Gamma\in\mathbb{R}^{+}$ and $K_{CL}\in\mathbb{R^{+}}$
are suitable observer gains. Since $d_{j}=\dot{\bar{s}}_{j}-\dot{s}_{j}$,
the approximated state derivative term $\dot{\bar{s}}_{j}$ is substituted
as $\dot{\bar{s}}_{j}=f_{m}(s_{j},\omega_{j})+\Omega^{T}(s_{j},v_{j})\chi_{j}+d_{j}$
to compute the estimation error dynamics. Using the observer equations
in (\ref{eq:StateEst})-(\ref{eq:DepthEst}), adding and subtracting
$K_{CL}\Gamma\sum_{j=1}^{M}\Omega(s_{j},v_{j})\Omega^{T}(s_{j},v_{j})\chi$,
grouping $\chi$ and $\hat{\chi}$, the estimation error dynamics
can be written as 
\begin{align}
\dot{\xi}= & -H\xi+\Omega^{T}(s,v)z\nonumber \\
\dot{z}= & -\Gamma\Omega(s,v)\xi+g(s,z,u)-K_{CL}\Gamma\big(\sum_{j=1}^{M}\Omega(s_{j},v_{j})d_{j}\nonumber \\
 & +\sum_{j=1}^{M}\Omega(s_{j},v_{j})\Omega^{T}(s_{j},v_{j})\left(z+\chi_{j}-\chi\right)\big)\label{eq:ErrorDyn2-2}
\end{align}
where $g(s,z,u)=f_{u}(s,\chi,u)-f_{u}(s,\hat{\chi},u)$.

\textbf{Assumption }$\textbf{4}$:\label{Assumption-:assumption2}
The history stack contains recent information and the change in depth
over a short period of time remains bounded i.e., $\exists\bar{\chi}\geq0$
such that $\sup_{t\geq0}\mathrm{max}_{j\in\left\{ 1,\cdots,M-1\right\} }\Vert\chi_{j}-\chi\Vert\leq\bar{\chi}$,
where $\chi_{j}=\chi\left(t_{j}\right)$ and $\chi$ are the past
and current true depth values at the time instants $t_{j}$ and $t$
such that $t>t_{j}\quad\forall j=1,\cdots,M-1$ for a suitably chosen
value of $M$.

\begin{remark}The main implication of Assumption 4 is that the history
stack should be frequently updated to contain information about the
current true depth from the past feature point and camera motion data.
Additionally, the upper bound $\bar{\chi}$ will be smaller if the
object is not too close to the camera and the camera linear velocities
are slow.\end{remark}

\section{Stability Analysis for Full Order Observer\label{sec:Stability-Analysis}}

Since the history stack is initialized with zeros, the stability analysis
is carried out in two phases, viz., the initial phase when the data
is being collected in the history stack and the phase when the history
stack is fully populated with informative points. In Theorem 1, leveraging
our prior work in \cite{dani2012globally}, it is shown that the estimation
error dynamics in (\ref{eq:ErrorDyn2-2}) are stable and yield a UUB
error under a PE condition when the history stack is incomplete. In
Theorem 2, it is shown that the estimation error dynamics in (\ref{eq:ErrorDyn2-2})
yield UUB error when the PE condition is not satisfied and the history
stack is complete. The advantage of adding the CL term is that the
error is bounded even if the PE condition is not satisfied. To facilitate
the analysis, let $\exists\bar{\sigma}>0$ such that $\sup_{t\geq0}\max_{j\in\left\{ 1,\cdots,M\right\} }\Vert\Omega(s_{j},v_{j})\Omega^{T}(s_{j},v_{j})\Vert\leq\bar{\sigma}$
and $\text{\ensuremath{\sigma_{1}}}=\sum_{j=1}^{M-1}\Omega(s_{j},v_{j})\Omega^{T}(s_{j},v_{j})$
such that $\sigma_{1}\in\mathbb{R}_{\geq0}$.

\begin{definition}The history stack is defined to be incomplete when
the stack is not completely populated with informative points such
that $\sigma_{1}\geq0.$ \end{definition}

\begin{definition}The history stack is defined to be complete when
the history stack is completely populated with informative points
such that $\sigma_{1}>0.$\end{definition} 
\begin{thm}
When the history stack is incomplete, the error system in (\ref{eq:ErrorDyn2-2})
is UUB if Assumption 4 and the PE condition in (\ref{eq:PECond})
are satisfied. Further, the ultimate bound on the estimation error
is given by $\sqrt{\frac{c_{2}}{c_{1}}}\frac{\gamma_{1}K_{CL}\Gamma((M-1)\bar{\sigma}\bar{\chi}+M\bar{d}\sqrt{\bar{\sigma}})}{k_{2}}$,
where $c_{1},c_{2},k_{2},\gamma_{1}$ are positive constants. 
\end{thm}
\begin{IEEEproof}
Refer Appendix $A$.
\end{IEEEproof}
\begin{thm}
When the history stack is complete, the error system in (\ref{eq:ErrorDyn2-2})
is UUB if Assumption 4 is satisfied, the PE condition in (\ref{eq:PECond})
is not satisfied, and the adjustable observer gain is selected according
to the sufficient condition, $K_{CL}>\frac{L_{g}}{\sigma_{1}\Gamma}$.
Further, the ultimate bound on the estimation error is given by $\sqrt{\frac{c_{4}}{c_{3}}}\frac{K_{CL}((M-1)\bar{\sigma}\bar{\chi}+M\bar{d}\sqrt{\bar{\sigma}})}{\sqrt{2k_{3}\alpha_{1}}}$,
where $c_{3},c_{4},k_{3},\alpha_{1}$ are positive constants.
\end{thm}
\begin{IEEEproof}
Refer Appendix $B$.
\end{IEEEproof}

\section{CL-based Reduced Order Observer\label{sec:Reduced-Order-Observer}}

For the reduced order CL observer, the history stack is a tuple $\mathcal{H=}\left\{ \left(s_{j},u_{j},\dot{v}_{j}\right)\right\} _{j=1}^{M}$
containing the past data points up to the index $M-1$ chosen by the
algorithm detailed in Section \ref{sec:History-Stack-Update} where
$M$ is the index of the data point at the current time instant. The
reduced order depth observer is defined as 
\begin{equation}
\hat{\chi}\left(t\right)=\kappa\left(s,\hat{\chi},u,\dot{v}\right)+\gamma\left(s,v\right)\label{eq:FinalEst}
\end{equation}
where 
\begin{align}
\dot{\kappa} & =f_{u}(s,\hat{\chi},u)+\bar{K}\sum_{j=1}^{M}\biggl(\theta_{j}^{T}\dot{v}_{j}\nonumber \\
 & \quad-\Omega(s_{j},v_{j})\left(f_{m}(s_{j},\omega_{j})+\Omega^{T}(s_{j},v_{j})\hat{\chi}\right)\biggr)\nonumber \\
\gamma & =-\bar{K}\sum_{j=1}^{M}\theta_{j}^{T}v_{j}\label{eq:kappa_gamma}
\end{align}
where $\theta_{j}=\left[x_{j}\;y_{j}\;\frac{-(x_{j}^{2}+y_{j}^{2})}{2}\right]^{T}$
for $j=1,\cdots,M$. The initial condition of the observer is selected
as $\kappa(t_{0})=\kappa_{0}$ where $\kappa_{0}>0$ is a constant.

\section{Stability Analysis for Reduced Order Observer\label{sec:Stability-Analysis-Reduced}}

Differentiating (\ref{eq:FinalEst}) and using (\ref{eq:kappa_gamma}),
to obtain the dynamics of $\hat{\chi}(t)$ as 
\begin{align}
\dot{\hat{\chi}}\negthickspace & =\negthickspace f_{u}\left(s,\hat{\chi},u\right)\negthickspace+\negthickspace\bar{K}\negmedspace\sum_{j=1}^{M}\negthickspace\Omega\left(s_{j},v_{j}\right)\negthinspace\left(\negthinspace\dot{s}_{j}\negthickspace-\negthickspace f_{m}(s_{j},\omega_{j})\negthickspace-\negthickspace\Omega^{T}(s_{j},v_{j})\hat{\chi}\right).\negthickspace\label{eq:RedCLObserver}
\end{align}
The error dynamics for the reduced order observer can be derived by
using (\ref{eq:Dynamics}), (\ref{eq:RedCLObserver}), substituting
$\dot{s}_{j}$ from (\ref{eq:SysDyn}), adding and subtracting $\bar{K}\sum_{j=1}^{M}\Omega(s_{j},v_{j})\Omega^{T}(s_{j},v_{j})\chi$,
and grouping $\chi$ and $\hat{\chi}$ as 
\begin{align}
\dot{z}\negthickspace & =-\bar{K}\negthickspace\left(\sum_{j=1}^{M}\negthickspace\Omega(s_{j},v_{j})\Omega^{T}(s_{j},v_{j})\left(z\negthickspace+\negthickspace\chi_{j}\negthickspace-\negthickspace\chi\right)\negthickspace\right)\negthickspace+g(s,z,u),\label{eq:FinalErrDyn}
\end{align}
where $g(s,z,u)=f_{u}(s,\chi,u)-f_{u}(s,\hat{\chi},u)$. Similar to
the case of the full order CL-based observer, the stability analysis
of the reduced order CL-based observer is carried out in two phases
viz. the initial phase when the data is being collected in the history
stack and the phase when the history stack is fully populated with
informative points. In Theorem 3, it is shown that the estimation
error dynamics in (\ref{eq:FinalErrDyn}) are stable and yield a UUB
error under a PE condition when the history stack is incomplete. In
Theorem 4, it is shown that the estimation error dynamics in (\ref{eq:FinalErrDyn})
yield UUB error when the PE condition is not satisfied and the history
stack is complete. 
\begin{thm}
When the history stack is incomplete, the error system in (\ref{eq:FinalErrDyn})
is UUB if Assumption 4 and the PE condition in (\ref{eq:PECond})
are satisfied. Further, the ultimate bound on the estimation error
is given by $\sqrt{\frac{c_{6}}{c_{5}}}\frac{\gamma_{2}\bar{K}\bar{\sigma}(M-1)\bar{\chi}}{k_{4}}$,
where $c_{5},c_{6},k_{4},\gamma_{2}$ are positive constants.
\end{thm}
\begin{IEEEproof}
Refer Appendix $C$.
\end{IEEEproof}
\begin{thm}
When the history stack is complete, the error system in (\ref{eq:FinalErrDyn})
is UUB if Assumption 4 is satisfied, the PE condition in (\ref{eq:PECond})
is not satisfied, and the adjustable observer gain is selected according
to the sufficient condition, $\bar{K}>\frac{L_{g}}{\sigma_{1}}$.
Further, the ultimate bound on the estimation error is given by $\frac{\bar{K}\bar{\sigma}(M-1)\bar{\chi}}{k_{5}}$,
where $k_{5}$ is a positive constants. 
\end{thm}
\begin{IEEEproof}
Refer Appendix $D$.
\end{IEEEproof}
\begin{remark}The gains $\bar{K}$ and $K_{CL}$ can be chosen to
minimize the effect of $z^{T}g(s,z,u)$. For the full order and the
reduced order observer, the estimation error decreases exponentially
to an ultimate bound as $t\to\infty$. The ultimate bound on the estimation
error can be made arbitrarily small by selecting appropriate gain
values $H$, $\Gamma$, $K_{CL}$ for the full order observer, or
$\bar{K}$ for the reduced order observer, and the size of the history
stack $M$. The optimal observer gains may be efficiently computed
by solving a Linear Matrix Inequality using incremental quadratic
constraints as demonstrated in \cite{chakrabarty2017state}.\end{remark}

\begin{remark}Old data can be replaced with new data in the history
stack even after the history stack is full as long as $\sigma_{1}$
is greater than zero. Using the procedure in Section \ref{sec:History-Stack-Update},
$\text{\ensuremath{\sigma_{1}}}$ always stays positive even after
the old points are replaced from the full history stack. Hence, the
upper bound on the derivative of the Lyapunov functions for the full
and reduced order observers holds at any given time after the history
stack is full. Thus, the ultimate bound on the switched systems can
be derived from analysis of multiple Lyapunov functions as demonstrated
in Thm 3.1 of \cite{vu2007input}.\end{remark}

\begin{remark}The ultimate bound on the estimation error $\iota_{b}$
increases linearly with $\bar{\chi}$, defined in Assumption 4. As
a result, if the history stack contains old points with previous values,
the ultimate bound on the estimation error will grow linearly with
$\bar{\chi}$. Hence, it is essential for the history stack to be
updated frequently to avoid the growth of the ultimate bound. Based
on the presented analysis, Algorithm I is designed to frequently update
the history stack and ensure that $\sigma_{1}$ stays greater than
zero.\end{remark}

\section{History Stack Update\label{sec:History-Stack-Update}}

From the analysis in Sections \ref{sec:Stability-Analysis} and \ref{sec:Stability-Analysis-Reduced},
an algorithm is designed in this section to ensure the ultimate bound
on the error is small and that $\sigma_{1}$ stays positive. Auxiliary
stacks $\mathcal{G=}\{(\dot{\bar{s}}_{j,}s_{j},u_{j})\}_{j=1}^{N}$
for the full order observer and $\mathcal{G=}\{(s_{j},u_{j},\dot{v}_{j})\}_{j=1}^{N}$
for the reduced order observer, such that $N>M$, are used to select
informative points. The auxiliary stack is a dynamic sliding window
of the $N$ most recent points. In each iteration, $M-1$ most informative
points are selected from the auxiliary stack to replace all the points
in the history stack. The history stack and auxiliary stack are both
initialized with zeros. At each time instance, the auxiliary stack
is sorted in descending order based on the value of $\Omega(s_{j},v_{j})\Omega^{T}(s_{j},v_{j})\quad\forall j=1,\cdots,N$
to select the top $M-1$ points. The points in the history stack are
replaced only if the chosen $M-1$ points from the auxiliary stack
satisfy $\ensuremath{\{\sum_{j=1}^{M-1}\Omega\Omega^{T}\}_{\mathcal{G}}\geq\epsilon}$
for a suitably chosen constant $\epsilon>0$. This, ensures that the
value of $\sigma_{1}$ does not drop below $\epsilon$ and the upper
bound on the derivative of the Lyapunov functions in (\ref{eq:DerivInequal})
and (\ref{eq:DerivInequal2-1}) holds at all times after the history
stack is full. The choice of $\epsilon$ is critical as it maintains
a balance between frequently updating the stack and ensuring that
$\sigma_{1}$ remains greater than zero. 
\begin{algorithm}[t]
\If{data is available}{ \If{${\mathcal{H}}$ is not full}{
Add data point to History Stack ${\mathcal{H}}$\; } Add data point
to ${\mathcal{G}}$ in a cyclic way\; \If{${\mathcal{H}}$ is full}{
Search for ${M-1}$ data points with maximum ${\{\Omega\Omega^{T}\}}$
in the ${\mathcal{G}}$ stack\; \If {${\{\sum_{j=1}^{M-1}\Omega\Omega^{T}\}}_{\mathcal{G}}\geq\epsilon$}
{ ${\mathcal{H}\leftarrow\mathcal{G}}$ for the selected ${M-1}$
points\; } } } \caption{Algorithm to update History Stack}
\label{I-Algorithm} 
\end{algorithm}

\section{Simulation Results\label{sec:Experiments}}

\subsection{Simulation 1\label{subsec:Simulation-1}}

A simulation is performed using a simulated feature point to verify
the performance of the CL full order observer designed in Section
\ref{sec:Concurrent-Learning-based-Observ}. An initial point with
Euclidean coordinates $\bar{m}(t_{0})=\left[2.5\:0.5\:3\right]{}^{T}$
is selected. A fourth order Runge-Kutta (R-K) ODE solver with a fixed
time-step of $\frac{1}{30}\mathrm{s}$ ($30\mathrm{fps}$) is used
to integrate the equations and generate trajectories for all values
of $\bar{m}(t)$. In this simulation, the linear velocities are designed
as $v=\left[0.3\:0.2\mathrm{cos}(\frac{\pi t}{4})\:-0.3\right]^{T}$and
$\omega=\left[0\:-\frac{\pi}{30}\:0\right]^{T}$. 
\begin{figure}[h]
\begin{centering}
\includegraphics[width=1\columnwidth]{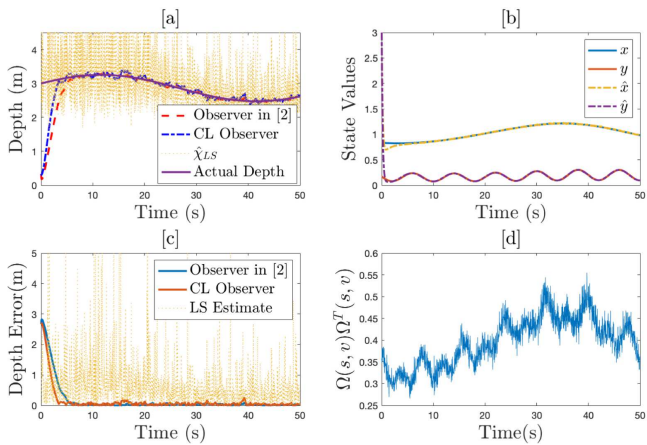} 
\par\end{centering}
\caption{(a) Comparison between the actual depth and estimated depth over a
period of 50 seconds.\label{fig:1(a)} (b) Comparison between the
actual and estimated state values.\label{fig:1(b)} (c) Error between
the true and the estimated depth. \label{fig:1(c)}(d) Evolution of
$\Omega(s,v)\Omega^{T}(s,v)$ for the simulated velocities and described
initial conditions. \label{fig:1(d)}}
\end{figure}

Additive white Gaussian noise with a signal to noise ratio (SNR) of
40 dB is added to the states and the velocity measurements are corrupted
with a Gaussian distributed measurement noise with zero mean and variance
of 0.01. The CL full order observer gain values used for the simulation
are $K_{CL}=0.15$, $\Gamma=5,$ $H=\mathrm{diag\{10,10\}}$. The
initial values for the state estimate and inverse depth estimate are
selected as $\hat{s}(t_{0})=\left[10\:5\right]^{T}$ and $\hat{\chi}(t_{0})=3$
which corresponds to a depth of $0.33\mathrm{m}$. The history stack
and the auxiliary stack are initialized with three points and five
points respectively. The CL full order observer converges to the true
depth in $4.7\:\mathrm{s}$ as shown in Figure \ref{fig:1(a)}(a).
Figure \ref{fig:1(b)}(b) shows the actual state trajectories and
the estimated state trajectories estimated by the CL full order observer.
The yellow dashed lines in Figure \ref{fig:1(a)}(a) show the performance
of the least squares (LS) depth estimation based on the formula $\hat{\chi}_{LS}=(\Omega^{T}(s,v))^{\dagger}(\dot{\bar{s}}-f_{m}(s,\omega))$.
Figure \ref{fig:1(c)}(c) shows the corresponding error plots for
the CL observer, the batch LS estimator and the observer in \cite{spica2014active}.
A simple least squares estimation is not a good solution to the depth
estimation problem due to the measurement noise and the singular value
of $\Omega$. The accuracy for the CL observer, batch LS estimator
and the observer in \cite{spica2014active} is reported for 500 Monte
Carlo runs of a 50s simulation. Initial conditions are sampled from
a normal distribution centered around $\hat{s}(t_{0})=[10,5]^{T}$
and $\hat{\chi}(t_{0})=3$. The CL full order observer achieves steady
state root mean square error (RMSE) of $0.046\mathrm{m}$ and mean
absolute percentage error (MAPE) of $1.83\%$. The LS estimation achieves
RMSE of $1.05\mathrm{m}$ and MAPE of $23.18\%$. The observer in
\cite{spica2014active} converges in $6.46\:\mathrm{s}$ and achieves
RMSE of $0.024\mathrm{m}$ and MAPE of $1.05\%$ when the gains are
set to $\mathbf{\Lambda}=9,\boldsymbol{H}=\mathrm{diag\{10,10\}},\boldsymbol{Q}=\mathrm{diag}\left\{ 4,4\right\} $.

\subsection{Simulation 2\label{subsec:Simulation-2}}

In this simulation the PE condition is violated between $31\mathrm{s}-\mathrm{38\mathrm{s}}$.
An initial point with Euclidean coordinates $\bar{m}(t_{0})=\begin{array}{c}
\left[1\:1\:1\right]^{T}\end{array}$ used to generate the trajectories using the velocities in simulation
1 from 0s to 31s. Since PE depends only on linear velocities, they
are chosen such that $\Omega(s,v)\Omega^{T}(s,v)=0$ at each time
instant during the period $31\mathrm{s}-38\mathrm{s}$. This implies
that $v_{X}=xv_{Z}$ and $v_{Y}=yv_{Z}$ at each time instant. The
linear velocity in the $Z$ direction is chosen to be $v_{Z}=0.1\mathrm{cos}(\frac{\pi t}{4})\mathrm{m/s}$
and the angular velocities are set to $0\:\mathrm{rad/s}$. As a result,
the linear and angular velocities are $v=\begin{array}{c}
\left[\frac{x(t)}{10}\mathrm{c}_{1},\;\frac{y(t)}{10}\mathrm{c}_{1},\;\frac{1}{10}\mathrm{c}_{1}\right]\end{array}^{T}$and $\omega=\left[0\:0\:0\right]^{T}$ where $\mathrm{c}_{1}=\mathrm{cos}(\frac{\pi t}{4})$.
For PE violation, the initial condition is set to the state at $31\mathrm{s}$
i.e $\bar{m}(t_{31})$. Once the PE violation stops at $38\mathrm{s}$,
the velocities in simulation 1 are used for simulating the trajectories
up to $50\mathrm{s}$. Additive white Gaussian noise with SNR $20\mathrm{dB}$
is added to the pixel measurements and Gaussian distributed noise
with zero mean and variance 0.01$\frac{\mathrm{m^{2}}}{\mathrm{s^{2}}}$
for the linear velocities and 0.01$\frac{\mathrm{rad}^{2}}{\mathrm{s^{2}}}$
for the angular velocities is added to the velocity measurements.
\begin{figure}[h]
\begin{centering}
\includegraphics[width=1\columnwidth]{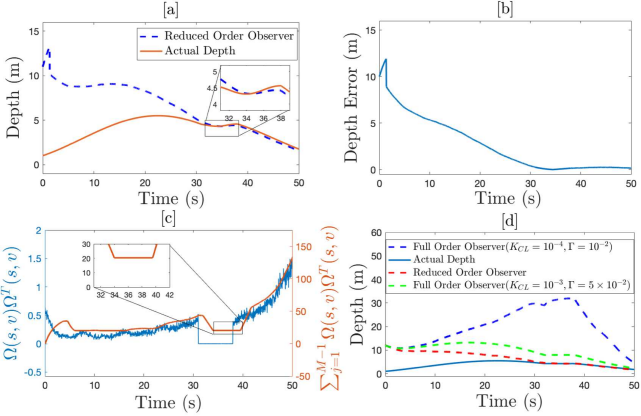} 
\par\end{centering}
\caption{(a) Comparison of true vs estimated depth using reduced order observer.
(b) Depth estimation error for reduced order observer. (c) Evolution
of $\Omega(s,v)\Omega^{T}(s,v)$ with time showing the PE violation
from $t=31\mathrm{s}$ to $38\mathrm{s}$. (d) Comparison between
reduced order integrated observer and CL observer for different gain
values in the presence of $20\mathrm{dB}$ noise.\label{fig:2}}
\end{figure}

The numerically approximated value of the PE between 31s to 38s is
$\intop_{31}^{38}\Omega(s(\tau),v(\tau))\Omega^{T}(s(\tau),v(\tau))d\tau=7.83\times10^{-4}$.
Figure \ref{fig:2}(a) shows the performance of the predicted depth
by the CL reduced order observer. The history stack is chosen to hold
$120$ points corresponding to $4\mathrm{s}$ of data and the auxiliary
stack is chosen to hold $150$ points corresponding to a window of
$5\mathrm{s}$. The observer is initialized at $\hat{s}(t_{0})=\left[1\;1\right]^{T}$
and $\hat{\chi}(t_{0})=0.08$ which corresponds to actual depth of
$12\mathrm{m}$. The gain $\bar{K}$ for the reduced order observer
is set to $2\times10^{-3}$. From Figure \ref{fig:2}(b) the CL reduced
order observer error exponentially converges to an ultimate bound
at $35.9\mathrm{s}$, during the period when the PE condition is violated.
The value of $\sigma_{1}=\sum_{j=1}^{M-1}\Omega(s_{j},v_{j})\Omega^{T}(s_{j},v_{j})$
always remains greater than zero once the history stack is full. The
value of $\sigma_{1}$ does not drop below $\epsilon$ defined in
Algorithm 1. The value of $\epsilon$ is chosen to be $20\mathrm{}$
and to maintain this value the history stack is not updated between
34s to 40s as shown in Figure \ref{fig:2}(c). The steady state RMSE
achieved is $0.129\mathrm{m}$ and steady state MAPE achieved is $3.61\%$.
Figure \ref{fig:2}(d) shows the comparison of the CL full order observer
presented in Section \ref{sec:Concurrent-Learning-based-Observ} with
the CL reduced order observer when the state measurements are noisy.

\section{Experiments}

\subsection{Experimental Platform}

The camera in the wrist of the right arm of a Baxter research robot
is used to capture images containing the feature point at a rate of
30 fps with resolution 640x400. The centroid of white circle is used
against a black background for easy thresholding based image segmentation.
The processing of the images and depth estimation is done in MATLAB
2019a at $30\:\mathrm{fps}$ using a desktop with Intel Core2Duo CPU
with clock-speed of 2.26 GHz and 4 GB RAM running Ubuntu 14.04. The
camera intrinsics for Baxter's right hand camera obtained through
the Baxter API and Robot Operating System (ROS) are given by $f_{x}=f_{y}=407.1$,
$c_{x}=323.4$ and $c_{y}=205.6$ where $\left(c_{x},c_{y}\right)$
represents the camera center pixel. The ground truth depth for comparing
the results is obtained using simple pose transformations as the pose
of the coordinate frame attached to camera and the feature point is
known in the coordinate frame attached to the base of the robot.

\subsection{Results}

\begin{figure}[h]
\begin{centering}
\includegraphics[width=1\columnwidth]{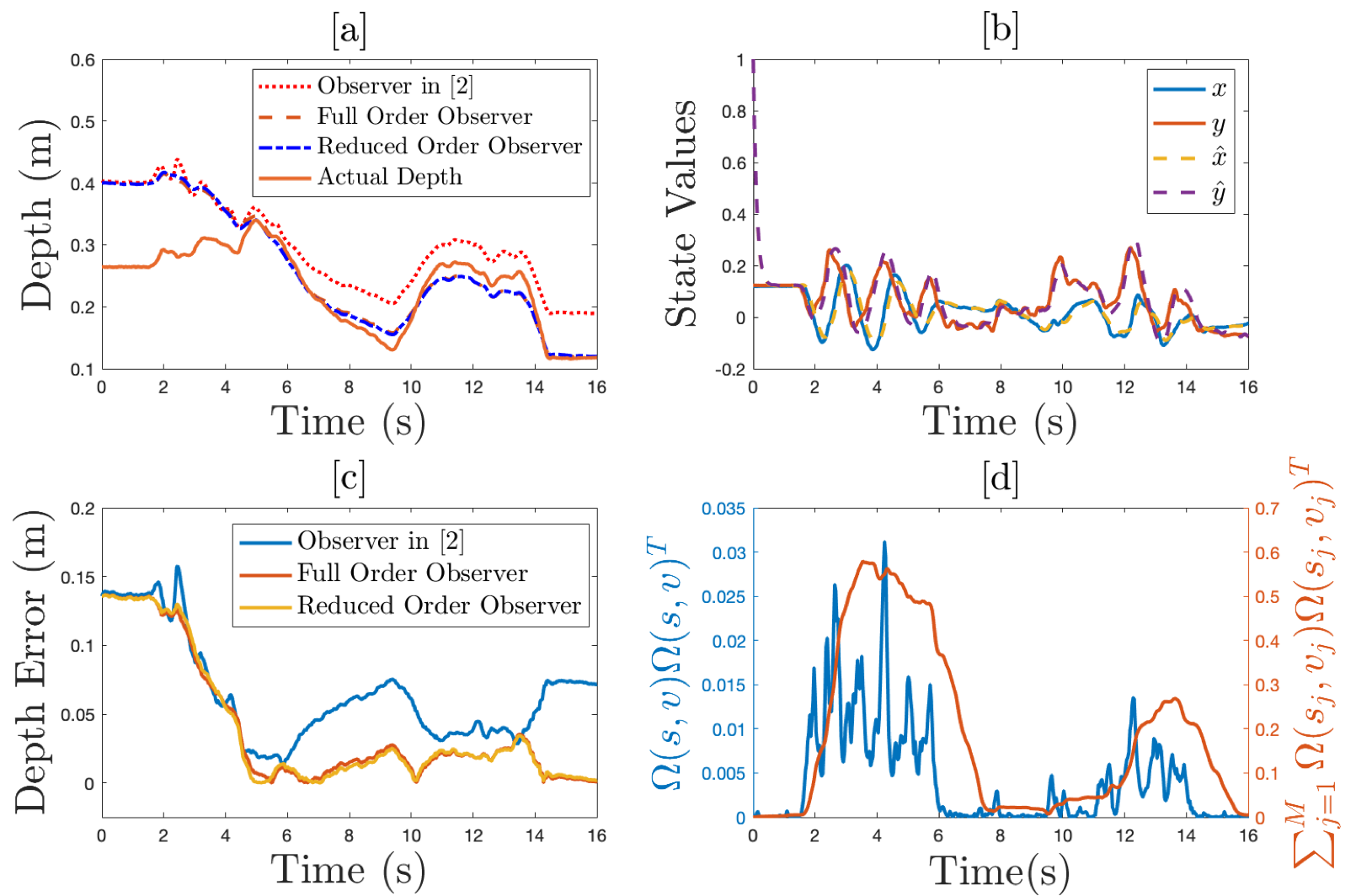} 
\par\end{centering}
\caption{(a) Comparison of true depth vs estimated depth using full order and
reduced order observer. (b) Comparison between the actual and estimated
state values using full order observer. (c) Depth estimation error
for full and reduced order observer. (d) Evolution of $\Omega(s,v)\Omega^{T}(s,v)$
and $\sum_{j=1}^{M}\Omega(s_{j},v_{j})\Omega^{T}(s_{j},v_{j})$ with
time.\label{fig:3}}
\end{figure}

The experiment is done for 16 seconds wherein the camera is stationary
for the first $1.5\mathrm{s}$. After $1.5\mathrm{s}$ the camera
starts moving in a circular motion in the $XY$ plane up to $5.5\mathrm{s}.$
After $5.5\mathrm{s}$ the camera moves downward along the $Z$ direction
and back for time up to $10\mathrm{s}.$ The motion from $10\mathrm{s}-15\mathrm{s}$
is circular in the $XY$ plane followed by a downward motion along
the $Z$ direction. The estimated depth by full and reduced order
observers, and the observer in \cite{spica2014active} is shown in
Figure \ref{fig:3}(a). The observers are initialized with initial
conditions $\hat{s}(t_{0})=\left[1\;1\right]^{T}$ and $\hat{\chi}(t_{0})=2.5$
which corresponds to a depth of $0.4\mathrm{m}$. 

The depth estimation error exponentially converges to an ultimate
bound as shown in Figure \ref{fig:3}(c). The camera moves in the
$XY$ plane from $1.5\mathrm{s}-5.5\mathrm{s}$ and the value of $\intop_{1.5}^{5.5}\Omega(s(\tau),v(\tau))\Omega^{T}(s(\tau),v(\tau))d\tau=4.12\times10^{-2}$
is approximated numerically using trapezoidal rule of integration.
When PE is violated $\intop_{6}^{10}\Omega(s(\tau),v(\tau))\Omega^{T}(s(\tau),v(\tau))d\tau=2.15\times10^{-3}\mathrm{}$,
also the maximum value of $\Omega(s,v)\Omega^{T}(s,v)$ is $0.035$
at $4.03\mathrm{s}$. The violation of the PE condition is achieved
by moving the camera along $Z$ direction when the feature point is
exactly at the centre of the image implying that $s\thickapprox0$.
The value of the estimated image plane feature point coordinates by
the observer in (\ref{eq:StateEst}) compared to the true values are
shown in Figure \ref{fig:3}(b). The value of the constant is chosen
as $\epsilon=0.03$ and $\sigma_{1}\geq\epsilon$ as shown in Figure
\ref{fig:3}(c) even when PE is not satisfied. The full order observer
converges in $4.7\mathrm{s}$ and achieves RMSE of $0.016\mathrm{m}$
and MAPE of $6.55\%$. The reduced order observer converges in $4.7\mathrm{s}$
and achieves RMSE of $0.015\mathrm{m}$ and MAPE of $6.28\%$. The
estimation error keeps decreasing in the first $6\mathrm{s}$ for
the observer in \cite{spica2014active} when the camera motion is
informative. However, the observer does not converge after $6\mathrm{s}$
when the motion is not informative. The observer in \cite{spica2014active}
achieves RMSE of $0.046\mathrm{m}$ and MAPE of $26.11\%$ when the
gains are set to $\mathbf{\Lambda}=60,\boldsymbol{H}=\mathrm{diag\{30,30\}},\boldsymbol{Q}=\mathrm{diag}\left\{ 6,6\right\} $.

\section{Conclusion\label{sec:Conclusion and Future Work}}

CL based full order and reduced order nonlinear observers are presented
in Section \ref{sec:Concurrent-Learning-based-Observ} and Section
\ref{sec:Reduced-Order-Observer} for estimating the depth of a stationary
feature point in an image using a moving camera. The estimation errors
for the full order observer and reduced order observer are shown to
be UUB. An analytical expression is derived for the ultimate bound
on the error for the presented observers. Based on the stability analyses,
an algorithm to update the history stack is designed in Section \ref{sec:History-Stack-Update}.
The algorithm ensures convergence of the error states and frequent
update of the history stack used for CL. The developed observers and
algorithm are verified through numerical simulations in Section \ref{subsec:Simulation-1}
and Section \ref{subsec:Simulation-2}. The observers are successfully
tested in real world experiments on a Baxter research robot when PE
is violated and the camera motions are not informative. Despite the
promising results, the choice of the sizes of the history and auxiliary
stacks, the observer gains, and $\epsilon$ are purely empirical.
The optimal choice of these parameters is a topic for the future.
The depth observer design for discrete time systems will also be explored
as a part of the future work.

 \bibliographystyle{IEEEtran}
\phantomsection\addcontentsline{toc}{section}{\refname}\bibliography{RCL_Complete}

\appendix
{\label{Appendix}}

\begin{lemma}\label{lemma_1}The function $g(s,z,u)$ is Lipschitz
continuous with respect to the variable $z$ with a Lipschitz constant
$L_{g}$.\end{lemma} 
\begin{IEEEproof}
For a single feature point , $g(s,z,u)=f_{u}(s,\chi,u)-f_{u}(s,\hat{\chi},u)$.
Using the definition of $f_{u}(s,\chi,u)$ in (\ref{eq:Dynamics})

\begin{align*}
\Vert g(s,z,u)\Vert & =\Vert f_{u}(s,\chi,u)-f_{u}(s,\hat{\chi},u)\Vert\\
 & =\Vert\chi^{2}v_{Z}+(y\omega_{X}-x\omega_{Y})\chi\\
 & \quad-(\hat{\chi}^{2}v_{Z}+(y\omega_{X}-x\omega_{Y})\hat{\chi})\Vert\\
 & =\Vert(\chi^{2}-\hat{\chi}^{2})v_{Z}+(y\omega_{X}-x\omega_{Y})(\chi-\hat{\chi})\Vert\\
 & =\Vert(\chi+\hat{\chi})v_{Z}z+(y\omega_{X}-x\omega_{Y})z\Vert\\
 & =\Vert\left((\chi+\hat{\chi})v_{Z}+(y\omega_{X}-x\omega_{Y})\right)z\Vert
\end{align*}
Using Assumption 2 and Remark 1, $g(\cdot)$ can be upper bounded
as 
\begin{align*}
\Vert g(s,z,u)\Vert & \leq\Vert(\chi+\hat{\chi})v_{Z}+(y\omega_{X}-x\omega_{Y})\Vert\Vert z\Vert\\
 & \leq L_{g}\Vert z\Vert
\end{align*}
The boundedness of $\hat{\chi}$ is ensured through the locally Lipschitz
projection law described in \cite{dani2012globally}. 
\end{IEEEproof}
\textit{A. Proof of Theorem 1}\\
Consider a domain $\mathcal{D}\subset\mathbb{R}^{3}$ containing $e(0)=[\xi(0),z(0)]^{T}$.
In the subsequent development, the result of Proposition 1 in \cite{DeLuca2008}
is used which proves the existence of a candidate Lyapunov function
$V(t,e):\mathcal{D}\rightarrow\mathbb{R}^{+}$ which can be upper
and lower bounded by $c_{1}\Vert e\Vert^{2}\leq V(t,e)\leq c_{2}\Vert e\Vert^{2}$
such that $c_{1},c_{2}>0$ and satisfies $\Vert\frac{\partial V}{\partial e}\Vert\leq\gamma_{1}\Vert e\Vert$.
The derivative of the Lyapunov function guarantees the exponential
stability of the estimation error dynamics without the CL terms when
the PE condition in (\ref{eq:PECond}) is satisfied. Using the triangle
inequality and Cauchy-Schwartz inequality an upper bound is derived
as $\sum_{j=1}^{M-1}\Omega(s_{j},v_{j})\Omega^{T}(s_{j},v_{j})(\chi_{j}-\chi)\leq\bar{\sigma}\sum_{j=1}^{M-1}\Vert\chi_{j}-\chi\Vert$.
Using the Lipschitz continuity property (refer Lemma 1), the term
$g(s,z,u)$ can be upper bounded by $\Vert g(s,z,u)\Vert\leq L_{g}\Vert z\Vert$,
where $L_{g}$ is the Lipschitz constant. Using the Cauchy-Schwartz
inequality and Lipschitz continuity of $g(s,z,u)$, the upper bounds
on the term $z^{T}g(s,z,u)$ can be derived as follows. 
\begin{equation}
\Vert z^{T}g(s,z,u)\Vert\leq L_{g}\Vert z\Vert{}^{2}\label{eq:inequal1}
\end{equation}
When the history stack is incomplete, $\sigma_{1}\geq0$. Using Assumption
3-4, result of Proposition 1 of \cite{DeLuca2008}, completing the
squares, the derivative of the Lyapunov function can be upper bounded
as 
\begin{align}
\dot{V} & \leq-\frac{k_{2}}{2c_{2}}V+\frac{\gamma_{1}^{2}K_{CL}^{2}\Gamma^{2}((M-1)\bar{\sigma}\bar{\chi}+M\bar{d}\sqrt{\bar{\sigma}})^{2}}{2k_{2}}\label{eq:VDot3-1}
\end{align}
for $k_{2}>0$. Using the comparison lemma 3.4 from \cite{Khalil2002},
the solution to the inequality in (\ref{eq:VDot3-1}) is given by
\begin{equation}
V(e(t))\leq V(e(t_{0}))e^{-\frac{k_{2}}{2c_{2}}\left(t-t_{0}\right)}+c_{2}\beta_{1}^{2}\left(1-e^{-\frac{k_{2}}{2c_{2}}\left(t-t_{0}\right)}\right)\label{eq:Lyapunov-Soln1}
\end{equation}
When the history stack is incomplete, the bound on the estimation
error $\Vert e(t)\Vert$ can be given as 
\begin{equation}
\Vert e(t)\Vert\leq\!\sqrt{\frac{c_{2}}{c_{1}}\left(\Vert e(t_{0})\Vert^{2}e^{-\frac{k_{2}}{2c_{2}}\left(t-t_{0}\right)}\!+\!\beta_{1}^{2}\left(1-e^{-\frac{k_{2}}{2c_{2}}\left(t-t_{0}\right)}\right)\right)}\label{eq:Error-Bound1}
\end{equation}
where $\beta_{1}=\frac{\gamma_{1}K_{CL}\Gamma((M-1)\bar{\sigma}\bar{\chi}+M\bar{d}\sqrt{\bar{\sigma}})}{k_{2}}$
yields an ultimate bound on estimation error $\left\Vert e(t)\right\Vert $
according to Theorem 4.18 of \cite{Khalil2002}. The error $\Vert e(t)\Vert$
is UUB with an ultimate bound $\iota_{b}=\sqrt{\frac{c_{2}}{c_{1}}}\beta_{1}$.

\textit{B. Proof of Theorem 2}\\
 Consider the candidate Lyapunov function $V(e):\mathcal{D}\rightarrow\mathbb{R}^{+}$
such that $V(e)=\frac{1}{2}\xi^{T}\xi+\frac{1}{2\Gamma}z^{T}z$ which
can be upper and lower bounded by constants $c_{3}\Vert e\Vert^{2}\leq V(e)\leq c_{4}\Vert e\Vert^{2}$
where $c_{3}=\mathrm{min}\left\{ \frac{1}{2},\frac{1}{2\Gamma}\right\} $
and $c_{4}=\mathrm{max}\left\{ \frac{1}{2},\frac{1}{2\Gamma}\right\} $.
The time derivative of the candidate Lyapunov function is considered
and the error dynamics in (\ref{eq:ErrorDyn2-2}) are used for analysis.
Since the history stack is complete, $\sum_{j=1}^{M-1}\Omega(s_{j},v_{j})\Omega^{T}(s_{j},v_{j})>0$,
PE is not satisfied, using (\ref{eq:inequal1}), completing the squares,
and considering the gain condition $K_{CL}>\frac{L_{g}}{\text{\ensuremath{\sigma_{1}\Gamma}}}$
is satisfied, the derivative of the Lyapunov function can be upper
bounded as
\begin{align}
\dot{V} & \leq-k_{1}\Vert\xi\Vert^{2}-\frac{k_{3}}{2}\Vert z\Vert^{2}+\frac{K_{CL}^{2}((M-1)\bar{\sigma}\bar{\chi}+M\bar{d}\sqrt{\bar{\sigma}})^{2}}{2k_{3}}\nonumber \\
 & \leq-\frac{\mathrm{min}\left\{ k_{1},\frac{k_{3}}{2}\right\} }{c_{4}}V+\frac{K_{CL}^{2}((M-1)\bar{\sigma}\bar{\chi}+M\bar{d}\sqrt{\bar{\sigma}})^{2}}{2k_{3}}\label{eq:DerivInequal}
\end{align}
such that $k_{1}=\lambda_{\mathrm{min}}\left\{ H\right\} $, $k_{3}=K_{CL}\text{\ensuremath{\sigma_{1}}}-\frac{L_{g}}{\Gamma}$,
$\alpha_{1}=\min\{k_{1},\frac{k_{3}}{2}\}$ where $\lambda_{\mathrm{min}}\left\{ \cdot\right\} $
is the minimum eigenvalue operator. Using the comparison lemma 3.4
from \cite{Khalil2002}, the solution to the inequality in (\ref{eq:VDot3-1})
is given by 
\begin{equation}
V(e(t))\leq V(e(t_{0}))e^{-\frac{\alpha_{1}}{c_{4}}\left(t-t_{0}\right)}+c_{4}\beta_{2}^{2}\left(1-e^{-\frac{\alpha_{1}}{c_{4}}\left(t-t_{0}\right)}\right)\label{eq:Lyapunov-Soln2}
\end{equation}
Subsequently, the bound on the estimation error $\Vert e(t)\Vert$
when the history stack is complete can be given as 
\begin{equation}
\Vert e(t)\Vert\leq\!\sqrt{\frac{c_{4}}{c_{3}}\left(\Vert e(t_{0})\Vert^{2}e^{-\frac{\alpha_{1}}{c_{4}}\left(t-t_{0}\right)}\!+\!\beta_{2}^{2}\left(1-e^{-\frac{\alpha_{1}}{c_{4}}\left(t-t_{0}\right)}\right)\right)}\label{eq:Error-Bound2}
\end{equation}
where $\beta_{2}=\frac{K_{CL}((M-1)\bar{\sigma}\bar{\chi}+M\bar{d}\sqrt{\bar{\sigma}})}{\sqrt{2k_{3}\alpha_{1}}}$.
Now, using the upper and lower bounds on $V(e)$, (\ref{eq:DerivInequal})
and invoking Theorem 4.18 in \cite{Khalil2002}, the error $\Vert e(t)\Vert$
is UUB with an ultimate bound $\iota_{b}=\sqrt{\frac{c_{4}}{c_{3}}}\beta_{2}$.

\textit{C. Proof of Theorem 3}\\
In the subsequent development, the result of Theorem 2 in \cite{dani2012globally}
is used which proves the existence of a candidate Lyapunov function
$V(t,z):[0,\infty)\times\mathbb{R}\rightarrow\mathbb{R}^{+}$ which
can be upper and lower bounded by $c_{5}\Vert z\Vert^{2}\leq V(t,z)\leq c_{6}\Vert z\Vert^{2}$
such that $c_{5},c_{6}>0$ and satisfies $\Vert\frac{\partial V}{\partial z}\Vert\leq\gamma_{2}\Vert z\Vert$.
The derivative of the Lyapunov function guarantees the exponential
stability of the estimation error dynamics without the CL terms when
the PE condition in (\ref{eq:PECond}) is satisfied. When the stack
is incomplete, $\sum_{j=1}^{M-1}\Omega(s_{j},v_{j})\Omega^{T}(s_{j},v_{j})\geq0$.
Using the dynamics in (\ref{eq:FinalErrDyn}), Assumption 4, definition
of the history stack, result of Theorem 2 in \cite{dani2012globally}
and completing the squares $\dot{V}$ can be upper bounded as 
\begin{align}
\dot{V} & \leq-\frac{k_{4}}{2}\Vert z\Vert^{2}+\frac{(\gamma_{2}\bar{K}\bar{\sigma}\left(M-1\right)\bar{\chi})^{2}}{2k_{4}}\nonumber \\
 & \leq-\frac{k_{4}}{2c_{6}}V+\frac{(\gamma_{2}\bar{K}\bar{\sigma}\left(M-1\right)\bar{\chi})^{2}}{2k_{4}}\label{eq:ReducedVdot3}
\end{align}
Using the comparison lemma 3.4 from \cite{Khalil2002}, the solution
to the inequality in (\ref{eq:VDot3-1}) is given by 
\begin{equation}
V\left(z\left(t\right)\right)\leq V\left(z\left(t_{0}\right)\right)e^{-\frac{k_{4}}{2c_{6}}\left(t-t_{0}\right)}+c_{6}\beta_{3}^{2}\left(1-e^{-\frac{k_{4}}{2c_{6}}\left(t-t_{0}\right)}\right)\label{eq:Lyapunov-Soln3}
\end{equation}
Subsequently using the development in Section 9.3 of \cite{Khalil2002},
the bound on the estimation error $\Vert z(t)\Vert$ when the history
stack is incomplete can be given as 
\begin{equation}
\Vert z(t)\Vert\leq\negmedspace\sqrt{\frac{c_{6}}{c_{5}}\left(\Vert z(t_{0})\Vert^{2}e^{-\frac{k_{4}}{2c_{6}}\left(t-t_{0}\right)}\negmedspace+\negmedspace\beta_{3}^{2}\left(1\negmedspace-\negmedspace e^{-\frac{k_{4}}{2c_{6}}\left(t-t_{0}\right)}\right)\negmedspace\right)}\label{eq:Error-Bound3}
\end{equation}
where $\beta_{3}=\frac{\gamma_{2}\bar{K}\bar{\sigma}\left(M-1\right)\bar{\chi}}{k_{4}}$,
yields an ultimate bound on estimation error $\Vert z(t)\Vert$ according
to Theorem 4.18 of \cite{Khalil2002}. The error $\Vert z(t)\Vert$
is UUB with an ultimate bound $\iota_{b}=\sqrt{\frac{c_{6}}{c_{5}}}\beta_{3}$.

\textit{D. Proof of Theorem 4}\\
 Consider the candidate Lyapunov function $V(z):\mathbb{R}\rightarrow\mathbb{R}^{+}$
such that $V=\frac{1}{2}z^{T}z$. When the history stack is complete
$\sum_{j=1}^{M-1}\Omega(s_{j},v_{j})\Omega^{T}(s_{j},v_{j})>0$, PE
is not satisfied, using (\ref{eq:inequal1}), considering the gain
condition $\bar{K}>\frac{L_{g}}{\text{\ensuremath{\sigma_{1}}}}$
to be satisfied and completing the squares, $\dot{V}$ can be upper
bounded as 
\begin{align}
\dot{V} & \leq-\frac{k_{5}}{2}\Vert z\Vert{}^{2}+\frac{(\bar{K}\bar{\sigma}\left(M-1\right)\bar{\chi})^{2}}{2k_{5}}\nonumber \\
 & \leq-k_{5}V+\frac{(\bar{K}\bar{\sigma}\left(M-1\right)\bar{\chi})^{2}}{2k_{5}}\label{eq:DerivInequal2-1}
\end{align}
Using the comparison lemma 3.4 from \cite{Khalil2002}, the solution
to the inequality in (\ref{eq:VDot3-1}) is given by 
\begin{equation}
V\left(z\left(t\right)\right)\leq V\left(z\left(t_{0}\right)\right)e^{-k_{5}\left(t-t_{0}\right)}+\frac{\beta_{4}^{2}}{2}\left(1-e^{-k_{5}\left(t-t_{0}\right)}\right)\label{eq:Lyapunov_Soln4}
\end{equation}
Subsequently, the bound on the estimation error $\Vert z(t)\Vert$
when the history stack is complete can be given as 
\begin{equation}
\Vert z(t)\Vert\leq\sqrt{\Vert z(t_{0})\Vert^{2}e^{-k_{5}\left(t-t_{0}\right)}+\beta_{4}^{2}\left(1-e^{-k_{5}\left(t-t_{0}\right)}\right)}\label{eq:Error-Bound4}
\end{equation}
where $k_{5}=(\bar{K}\text{\ensuremath{\sigma_{1}}}-L_{g})$ and $\beta_{4}=\frac{\bar{K}\bar{\sigma}\left(M-1\right)\bar{\chi}}{k_{5}}$.
Using (\ref{eq:DerivInequal2-1}) and invoking Theorem 4.18 in \cite{Khalil2002},
the depth error $\Vert z(t)\Vert$ is UUB with an ultimate bound $\iota_{b}=\beta_{4}$
. 
\end{document}